\documentclass[journal, draftclsnofoot,onecolumn,12pt]{IEEEtran}

\usepackage[T1]{fontenc}

\ifCLASSINFOpdf
\else
\fi
\usepackage{amsmath}
\usepackage[cmintegrals]{newtxmath}
\usepackage{ amssymb }
\usepackage{amsfonts}

\usepackage{graphicx}
\usepackage{lettrine}
\usepackage{epstopdf}
\usepackage{textcomp,booktabs}
\usepackage[usenames,dvipsnames]{color}
\usepackage{colortbl}
\definecolor{mygray}{gray}{.9}
\definecolor{mypink}{rgb}{.99,.91,.95}
\definecolor{mycyan}{cmyk}{.3,0,0,0}
\usepackage{pifont}
\usepackage{subfig}
\usepackage{multirow}

\usepackage{color}

\usepackage[dvipsnames,usenames]{color}
\usepackage[usenames,dvipsnames]{color}

\definecolor{light-gray}{gray}{0.90}
\begin{document}
%
\title{ComNet: Combination of Deep Learning and Expert Knowledge in OFDM Receivers}
%
%
%

\author{Xuanxuan~Gao,~\IEEEmembership{Student~Member,~IEEE,}
	    Shi~Jin,~\IEEEmembership{Senior~Member,~IEEE,}
        Chao-Kai~Wen,~\IEEEmembership{Member,~IEEE,}
        and Geoffrey Ye Li,~\IEEEmembership{Fellow,~IEEE}
\thanks{X. Gao and S. Jin are with the National Mobile Communications Research Laboratory, Southeast University, Nanjing 210096, China (e-mail: gaoxuanxuan@seu.edu.cn; jinshi@seu.edu.cn).}
\thanks{C.-K. Wen is with the Institute of Communications Engineering, National Sun Yat-sen University, Kaohsiung 804, Taiwan (e-mail: chaokai.wen@mail.nsysu.edu.tw).}
\thanks{G. Y. Li is with the School of Electrical and Computer Engineering, Georgia Institute of Technology, Atlanta, GA 30332 USA (email: liye@ece.gatech.edu).}}
\maketitle

\vspace{-0.5cm}
\begin{abstract}
In this article, we propose a model-driven deep learning (DL) approach that combines DL with the expert knowledge to replace the existing orthogonal frequency-division multiplexing (OFDM) receiver in wireless communications. Different from the data-driven fully connected deep neural network (FC-DNN) method, we adopt the block-by-block signal processing method that divides the receiver into channel estimation subnet and signal detection subnet. Each subnet is constructed by a DNN and uses the existing simple and traditional solution as initialization. The proposed model-driven DL receiver offers more accurate channel estimation comparing with the linear minimum mean-squared error (LMMSE) method and exhibits higher data recovery accuracy comparing with the existing methods and FC-DNN. Simulation results further demonstrate the robustness of the proposed approach in terms of signal-to-noise ratio and its superiority to the FC-DNN approach in the computational complexities or the memory usage.
\end{abstract}

\begin{IEEEkeywords}
Deep learning, wireless communications, OFDM.
\end{IEEEkeywords}

%
\IEEEpeerreviewmaketitle

\vspace{-0.5cm}
\section{Introduction}
%
%
%
%
Deep learning (DL) has gained great successes in the fields of computer vision and natural language processing, among others, and has been considered for application in wireless communications since then. The potential applications of DL in physical layers are discussed in \cite{8054694}, \cite{8233654} and \cite{qin2018deep}. The data-driven
DL approach in \cite{8052521} adopts a fully connected deep
neural network (FC-DNN) and replaces all modules at the
orthogonal frequency-division multiplexing (OFDM) receiver.
The abovementioned literature challenges the conventional
OFDM receiver and treats the receiver as a black box. But, it does not exploit the expert knowledge\footnote{The term ``expert knowledge'' used in this article means block-based architectures and algorithms in conventional wireless communications and differs from the knowledge-based expert system in the artificial intelligence field.} in wireless communications, which in turn renders the FC-DNN-based receiver unexplainable and unpredictable. In addition, the data-driven method relies on a huge amount of data to train a large number of parameters, thus converges slowly and has high computational complexity.

To address the above issues, the model-driven DL approach can be used instead. A general model-driven DL framework in \cite{doi:10.1093/nsr/nwx099} can overcome the heavy demand of a huge amount of training data. Moreover, the model-driven DL network can clearly explain the specially designed model family by using domain knowledge to facilitate further performance improvement. In the field of wireless communications, all modules in transceivers have been rigorously developed, which subsequently enables to set the existing algorithms as the fundamentals of the model family in model-driven DL approaches. {The superiority of introducing expert knowledge into wireless communications to form a model-driven DL solution has been demonstrated in the examples of the radio transformer network (RTN) \cite{8054694}, the channel state information (CSI)-aided MIMO communication \cite{modeldrivenmethod} and the PAPR reducing network (PRNet) \cite{kim2018novel}.} {A comprehensive overview of model-driven DL for physical layer communications can be found in \cite{he2018modeldriven}.}



In this article, we propose a model-driven DL {architecture}, called ComNet, to replace the conventional or FC-DNN OFDM receiver \cite{8052521}, which combines DL with expert knowledge in wireless communications.
{The proposed ComNet receiver uses DL to facilitate existing receiver models, such as channel estimation (CE) module and signal detection (SD) module, rather than replacing the receiver with an entire DL architecture and then integrating communication information, such as the RTN \cite{8054694}, CSI-aided MIMO communications \cite{modeldrivenmethod} and PRNet \cite{kim2018novel}.} This model-driven DL approach shows better performance comparing with the traditional LMMSE-MMSE method and FC-DNN \cite{8052521} and exhibits relatively faster convergence speed with fewer parameters comparing with the FC-DNN OFDM receiver \cite{8052521}.

\vspace{-0.2cm}
\section{ComNet}
This section presents the proposed ComNet receiver {for the OFDM system}. The architecture and the details of {the DL-based subnets, including the CE and SD subnets,} are elaborated in Section \uppercase\expandafter{\romannumeral2} A. In Section \uppercase\expandafter{\romannumeral2} B, initializations of network weights, the choice of cost function and optimizer, and configurations of hyper-parameters are explained.

\vspace{-0.3cm}
\subsection{ComNet Architecture}
\begin{figure}[!t]
	\centering
	\includegraphics[width=5in]{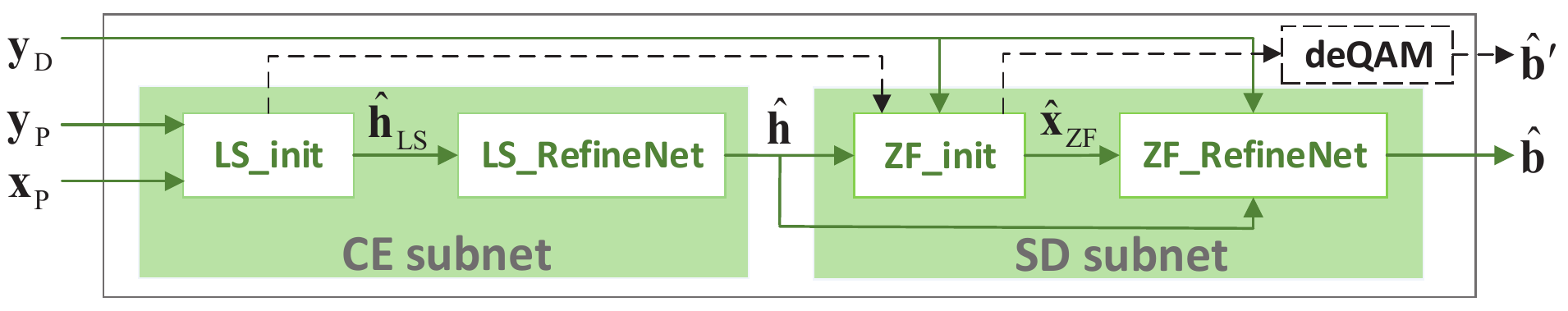}
	\caption{ComNet receiver architecture. The two subnets use traditional communication solutions as initializations, and apply DL networks to refine the coarse inputs. The dotted short-path provides a relatively robust candidate of the binary symbols recovery.}
	\label{ComNet}
\end{figure}

{In the OFDM system, the transmitted signals consist of the transmitted data vector, ${{\bf x}_{\rm D}}$, and the pilot symbol vector, ${{\bf x}_{\rm P}}$, which is known to the receiver. Correspondingly, the received signals include the received data vector, ${{\bf y}_{\rm D}}$, and the received pilot symbol vector, ${{\bf y}_{\rm P}}$. The conventional OFDM receiver recovers the estimation of the transmitted binary data ${\hat{\bf b}}$ given frequency domain signals, ${{\bf y}_{\rm D}}$, ${{\bf y}_{\rm P}}$, and ${{\bf x}_{\rm P}}$ through CE, SD, and quadrature amplitude modulation (QAM) demodulation sequentially.

Fig. \ref{ComNet} illustrates the architecture of the ComNet receiver. The inputs and outputs of the ComNet receiver are similar to that of the conventional OFDM receiver while the ComNet receiver adopts two cascaded DL-based subnets to replace the conventional OFDM receiver. Instead of using straightforward FC-DNN as in \cite{8052521} to estimate the transmitted data in a brute force manner, in the proposed ComNet receiver, the CE and SD subnets use traditional communication solutions as initializations and apply DL networks to refine the coarse inputs. The ComNet receiver also takes full advantage of the conventional methods and connects them to form a relatively robust recovery to adapt to various scenarios.}

\setlength{\belowcaptionskip}{-0.3cm}   

\setlength{\belowcaptionskip}{-0.3cm}   
\begin{figure}[!t]
	\centering
	\includegraphics[width=3in]{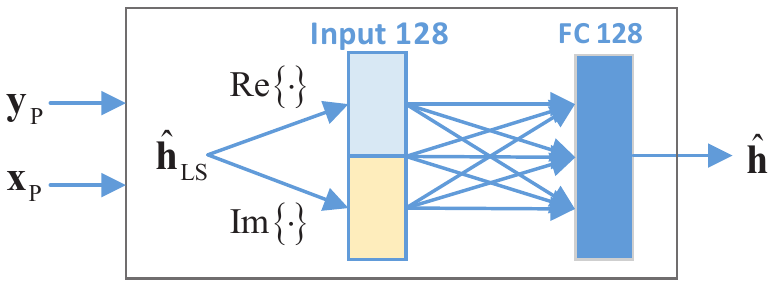}
	\caption{CE subnet. A subnet type initialized by LS CE. Then the real-valued initialization is refined by LS\_RefineNet.}
	\label{subnet1}
\end{figure}

\setlength{\belowcaptionskip}{-0.5cm}   
\begin{figure}[!t]
	\centering
	\includegraphics[width=4.2in]{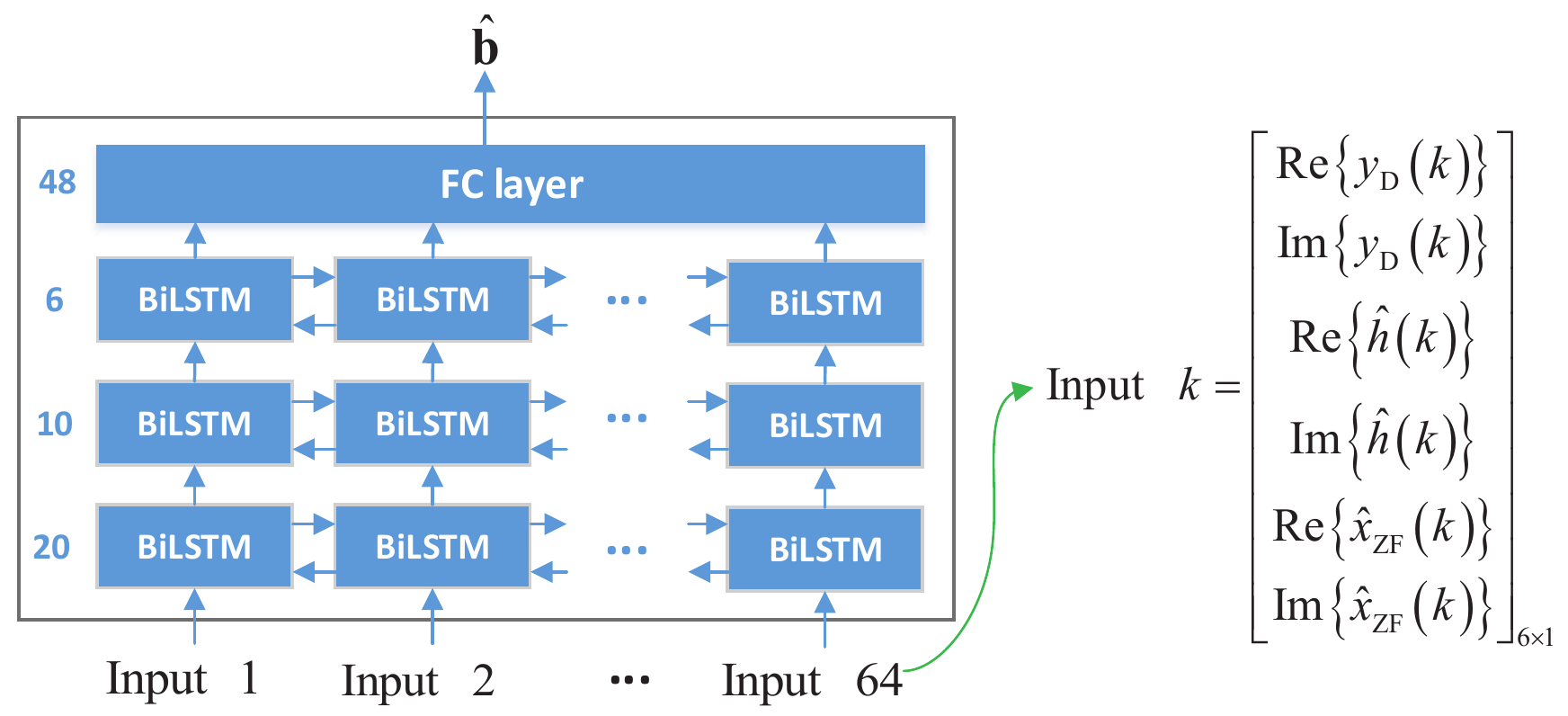}
	\caption{Detection subnet with BiLSTM-Detection as ZF\_RefineNet. A subnet type initialized by ZF solution, in which ZF\_RefineNet adopts BiLSTM-Detection with a FC-DNN layer.}
	\label{subnet2}
\end{figure}

Fig. \ref{subnet1} shows the CE subnet. Its input is the least-square (LS) CE, obtained by
\begin{equation}
{{\hat {h}}_{{\rm{LS}}}}(k) = \frac{{{{y}_{\rm P}}(k)}}{{{{{x}_{\rm P}}}(k)}},
\end{equation}
where ${{x}_{\rm P}(k)}$ and ${{y}_{\rm P}(k)}$ are the pilot symbol and the corresponding received symbol at the $k$-th subcarrier. Then ${{\hat{\bf h}}_{{\rm{LS}}}}$ is used by the LS\_RefineNet to generate accurate CE ${\hat{\bf h}}$, where the LS\_RefineNet is a one-layer DNN. Its input is a 128-dimensional real-valued signal vector that consists of the real and imaginary parts of ${{\hat{\bf h}}_{{\rm{LS}}}}$. The number of neurons of the following layer is 128 and these neurons have no activation function, that is, it is a linear channel estimator.

In the SD subnet, the input is just zero-forcing (ZF) SD of the transmit symbol, obtained by
\begin{equation}
{{\hat{x}}_{{\rm{ZF}}}}(k) = \frac{{{{y}_{\rm D}}(k)}}{{{\hat{h}}(k)}}.
\end{equation}
The ${{\hat{\bf x}}_{{\rm{ZF}}}}$ is used by the ZF\_RefineNet to predict the binary data from 8 symbols on 8 consecutive subcarriers. For an OFDM system with 64 subcarriers, 8 independent SD subnets are needed. In summary, the ZF\_RefineNet uses ${{\hat{\bf x}}_{{\rm{ZF}}}}$, ${\hat{\bf h}}$, and ${{\bf y}_{\rm D}}$ to get more accurate estimation of the transmit data. Depending on different requirements on receiver complexity and data recovery accuracy, we propose two different forms of ZF\_RefineNet.
\begin{itemize}
	\item \textbf{FC-SD} involves a two-layer FC-DNN with 120 and 48 neurons on each layer. The input is the concatenation of the real and imaginary parts of ${{\hat{\bf x}}_{{\rm{ZF}}}}$. The activation function of the hidden layer uses a ReLU function, $f_{\rm Re}(a)=\max ({0,a})$, whereas that of the output layer is the logistic sigmoid function, $f_{\rm Si}(a)=\frac{1}{{1 + {e^{ - a}}}}$.
	\item \textbf{Bi-directional long short-term memory (BiLSTM)-SD} involves a three-layer 64-time steps BiLSTM network \cite{1556215} with 20, 10 and 6 hidden units for each layer, followed by an one-layer FC-DNN with 48 neurons, as in Fig. \ref{subnet2}. {Considering the performance degradation of the ZF SD, the inputs of BiLSTM-SD integrate ${{\bf y}_{\rm D}}$ and ${\hat {\bf h}}$ as well.} The activation function of the output layer is the logistic sigmoid function as before.
\end{itemize}

The abovementioned 48 outputs correspond to 48 bits to be estimated from eight consecutive subcarriers with 6 bits for each symbol for 64-QAM. Because the logistic sigmoid function maps the input to the interval, $[0,1]$, the received binary symbol will be ``1'' if the output is larger than 0.5 and ``0'' otherwise. {However, the number of layers and that of neurons in each layer, except for the last layer, depend on the empirical trials.}

\vspace{-0.4cm}
\subsection{Training Specification}
\vspace{-0.1cm}
%
%
%


To accelerate the training process, the initialization of the network weights is considered. The CE subnet is initialized by the real-valued linear minimum mean-squared error (LMMSE) CE weight matrix ${{\tilde{\bf W}}_{{\rm{LMMSE}}}}$ from 
\begin{equation}
{{\tilde{\bf h}}_{{\rm{LMMSE}}}} = {{\tilde{\bf W}}_{{\rm{LMMSE}}}}{{\tilde{\bf h}}_{{\rm{LS}}}},
\end{equation}
where
\begin{equation}
{{\tilde{\bf W}}_{{\rm{LMMSE}}}}{\rm{ = }}\left[ {\begin{array}{*{20}{c}}
	{{\rm{Re}}\left\{ {{{\bf{W}}_{{\rm{LMMSE}}}}} \right\}}&{{\rm{ - Im}}\left\{ {{{\bf{W}}_{{\rm{LMMSE}}}}} \right\}}\\
	{{\rm{Im}}\left\{ {{{\bf{W}}_{{\rm{LMMSE}}}}} \right\}}&{{\rm{Re}}\left\{ {{{\bf{W}}_{{\rm{LMMSE}}}}} \right\}}
	\end{array}} \right].
\end{equation}
${{\tilde{\bf h}}_{{\rm{LMMSE}}}}$ and ${{\tilde{\bf h}}_{{\rm{LS}}}}$ are the concatenation of real and imaginary parts of LMMSE CE ${{\hat{\bf h}}_{{\rm{LMMSE}}}}$ and LS CE ${{\hat{\bf h}}_{{\rm{LS}}}}$, respectively. In particular, the LMMSE CE weight matrix, ${{\hat{\bf W}}_{{\rm{LMMSE}}}}$, adopts the method in \cite{Cho:2010:MWC:1951613}. The multiplicative weights in the FC layers of the SD subnet are initialized by the method in \cite{7410480}. 

%
%

After initialization, the DL networks are trained by minimizing the cost between ${\hat{\bf b}}$ and the raw transmit binary symbol ${\bf{b}}$ to adjust the network parameters. {The training data are obtained from simulation under system configurations in the next section.} We adopt the mean-squared error cost function as in \cite{8052521} and the adaptive moment estimation (Adam) optimizer \cite{Kingma2014Adam} for both subnets. 
{The two subnets are trained sequentially in TensorFlow, in which the CE subnet is trained for 2,000 epochs and then fixed, followed by 5,000 epochs training for the SD subnet. } 
The end-to-end comparison shows that sequential training can guarantee the optimality of each block and speed up the training process with fewer network parameter requirements. Each epoch utilizes 50 mini-batches for a total batch size of 1,000. The learning rate is set as a staircase function to realize training epochs with initial values of 0.001 and decreased 10-fold every 1,000 epochs for the CE subnet and decreased 5-fold every 2,000 epochs for the SD subnet.

\vspace{-0.3cm}
\section{Numerical Results}
Simulation is conducted under three cases as in \cite{8052521}, linear, cyclic prefix (CP) removal, and clipping, which are corresponding to no postfix, ``\_CP'' postfix and ``\_CR'' postfix in figures. {Since the main contribution of this article is to propose a novel model-driven OFDM receiver architecture instead of resolving the nonlinearity, traditional nonlinearity compensatory methods are not delved.} The ``SameSNR'' markers represent the results when the ComNet receiver is trained and deployed both under $\rm{SNR}=5~dB$. Simulation results are compared in terms of accuracy and complexity among the proposed ComNet receiver, FC-DNN receiver \cite{8052521} and the traditional communication methods \cite{Cho:2010:MWC:1951613}. 

System configurations of the simulation are similar to \cite{8052521} as follows. The OFDM system contains 64 subcarriers with 16 samples of CP and each frame contains one pilot OFDM symbol and one data OFDM symbol. The mapping of 64-QAM adopts the long-term evolution (LTE) standard. The channel is a WINNER \uppercase\expandafter{\romannumeral2} channel under C1 scenario NLOS case in 2.6 GHz. The ComNet receiver utilizes FC-SD in the linear case, whereas BiLSTM-SD is adopted in the nonlinear cases.

We employ the following concise conventions in the subsequent discussion:
\begin{itemize}
	\item \textbf{ComNet-BiLSTM}: Proposed ComNet architecture with BiLSTM-Detection ZF\_RefineNet
	\item \textbf{ComNet-FC}: Proposed ComNet architecture with FC-Detection ZF\_RefineNet
	\item \textbf{FC-DNN}: FC-DNN in \cite{8052521}, but modified such that the number of neurons in the output layer is changed from 16 to 48 to render it suitable for 64-QAM
	\item \textbf{LMMSE-MMSE}: Traditional LMMSE CE and minimum mean-squared error (MMSE) SD
	\item \textbf{Y/H\_true}: The quotient of ${{\bf y}_{\rm D}}$ and the true frequency domain channel ${\bf h}$, which can realize the maximum-likelihood solution in the linear case.
\end{itemize}

\vspace{-0.5cm}
\subsection{CE Subnet}
\begin{figure}[!t]
	\centering
	\includegraphics[width=3.5in]{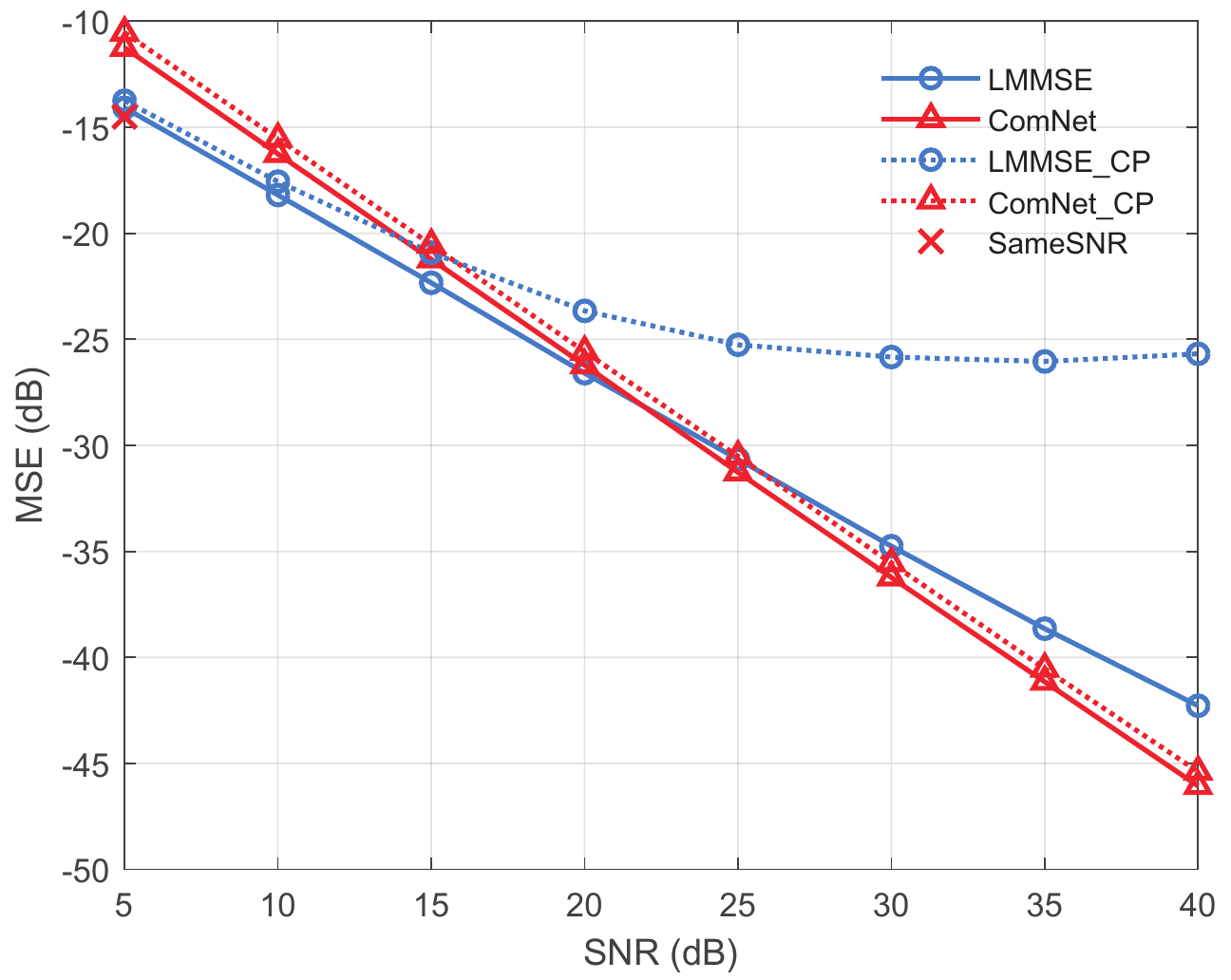}
	\caption{MSE curves of ComNet and traditional methods under linear case and CP removal case.}
	\label{MSE}
\end{figure}

A benefit of using ComNet receiver against FC-DNN is to access accurate CSI, which is useful for channel analysis and CSI feedback in downlink transmission. Fig. \ref{MSE} shows MSE performance of the ComNet CE subnet and the LMMSE method under linear and CP removal cases. From the figure, the CE subnet can better rectify the effect introduced by CP removal compared with the traditional LMMSE CE. This is as a result that the training process of LS\_RefineNet modifies the network multiplicative weights from the initialized value, ${{\tilde{\bf W}}_{{\rm{LMMSE}}}}$, to proper values through minimizing the channel MSE with the Adam optimizer.

\setlength{\belowcaptionskip}{-0.3cm}   

\vspace{-0.5cm}
\subsection{SD Subnet}
\subsubsection{Linear Case}
\begin{figure}[!t]
	\centering
	\includegraphics[width=3.5in]{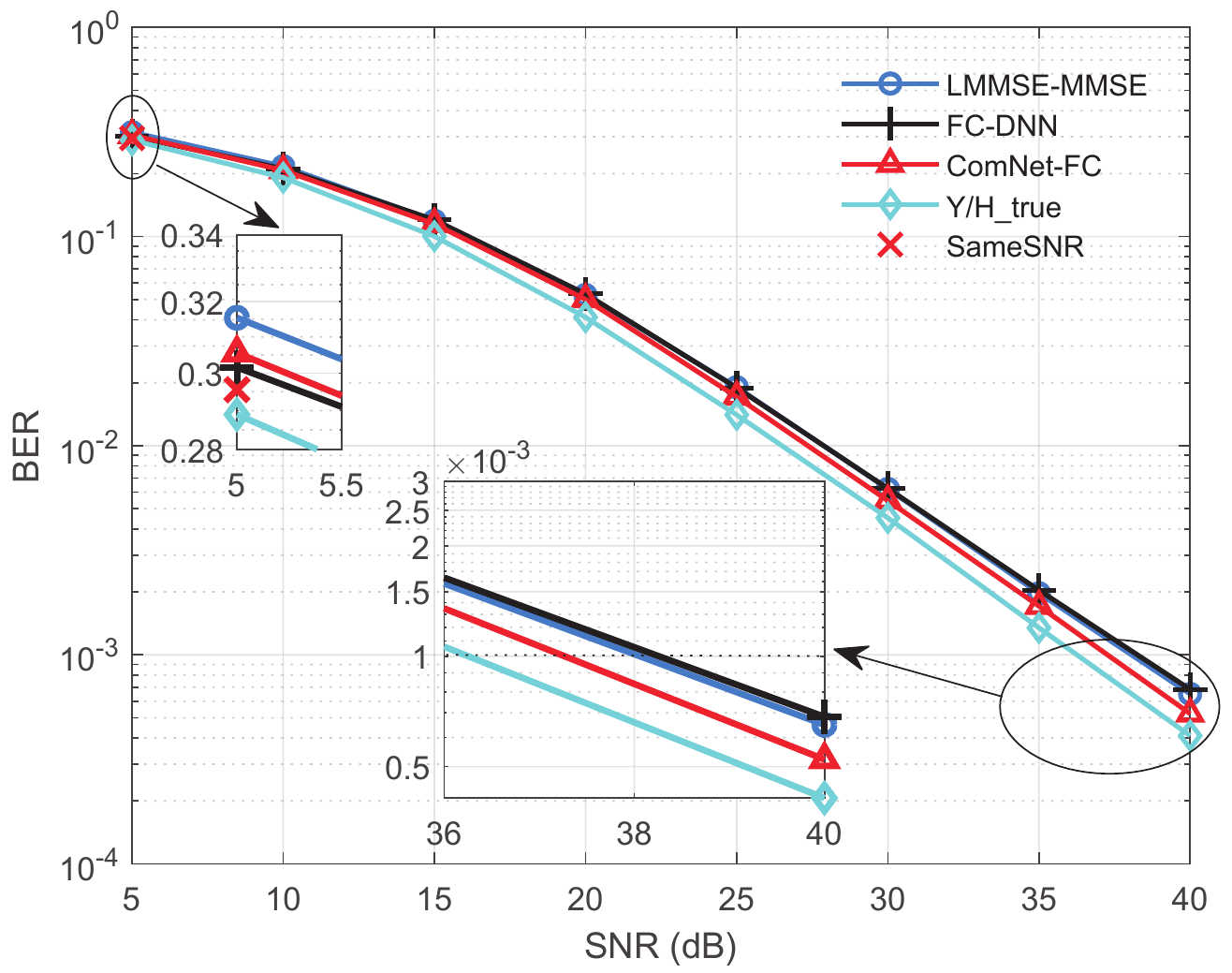}
	\caption{BER curves of ComNet and competing methods under linear case.}
	\label{Linear}
\end{figure}

Fig. \ref{Linear} compares the bit-error rate (BER) curves of ComNet-FC and existing methods under the linear case, where a basic OFDM system without nonlinear effects is considered. From the figure, the BER of the proposed ComNet receiver is closest to the ideal bound Y/H\_true compared with FC-DNN and LMMSE-MMSE. The required signal-to-noise ratio (SNR) for the ComNet receiver to reach $\rm{BER}=10^{-3}$ is 1 dB better than FC-DNN and LMMSE-MMSE. But there is also an 1 dB gap between the ComNet receiver and the ideal bound. {Extra simulation suggests that the ComNet receiver obviously outperforms the FC-DNN under longer delay spread.}

The model-driven approach, ComNet-FC receiver, has only one-eighth amount of parameters compared with the data-driven approach of FC-DNN \cite{8052521}. Furthermore, the ComNet-FC receiver needs only 200 epochs to converge while FC-DNN needs approximately 2,000 epochs to reach the same BER level. This result demonstrates the superior convergent speed and minimal parameter requirements of the model-driven DL approach.
\setlength{\belowcaptionskip}{-0.5cm}   
\setlength{\belowcaptionskip}{-0.1cm}   
\begin{figure}[!t]
	\centering
	\subfloat[ ]{
		\includegraphics[width=3.5in]{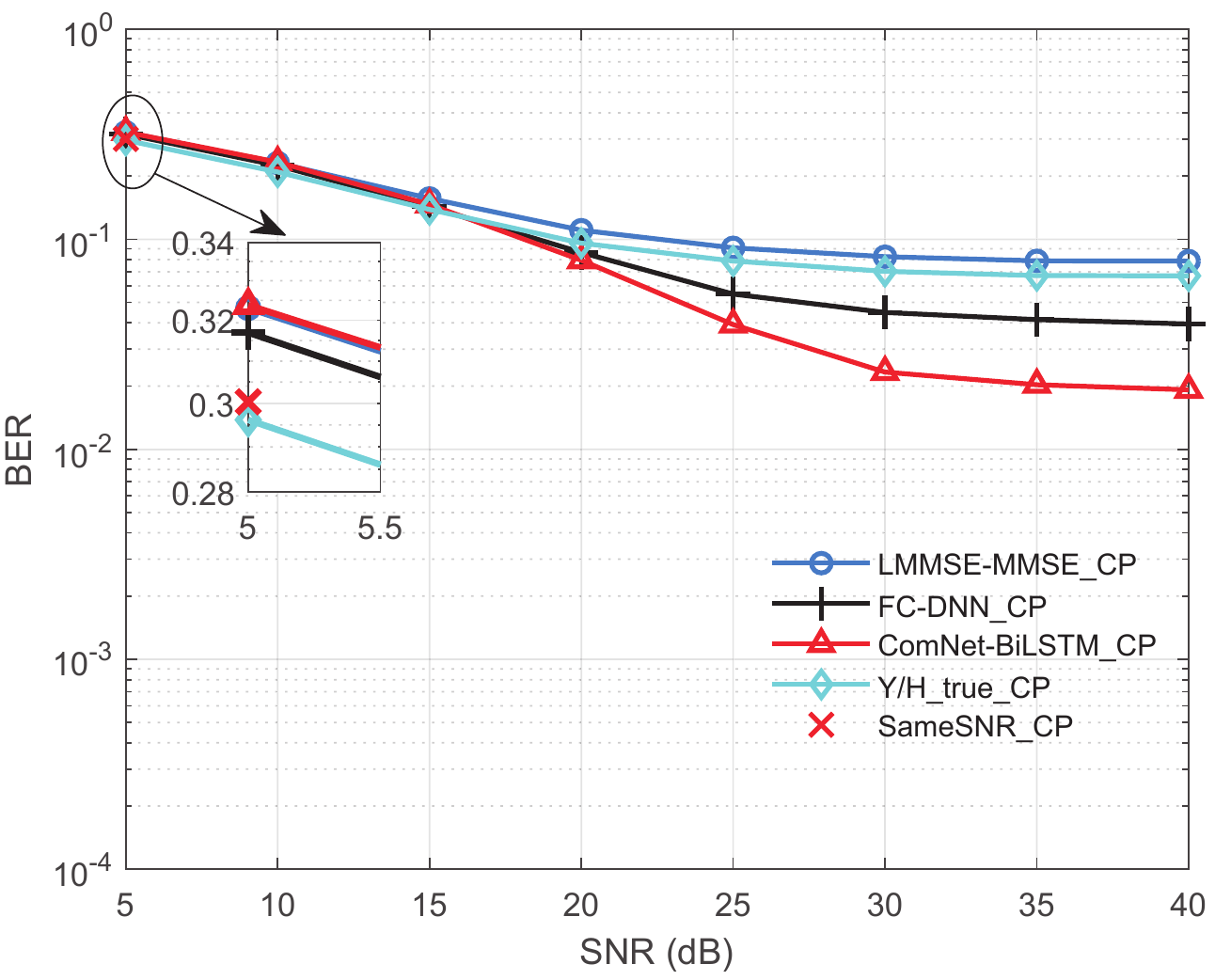}}\\
	\subfloat[ ]{
		\includegraphics[width=3.5in]{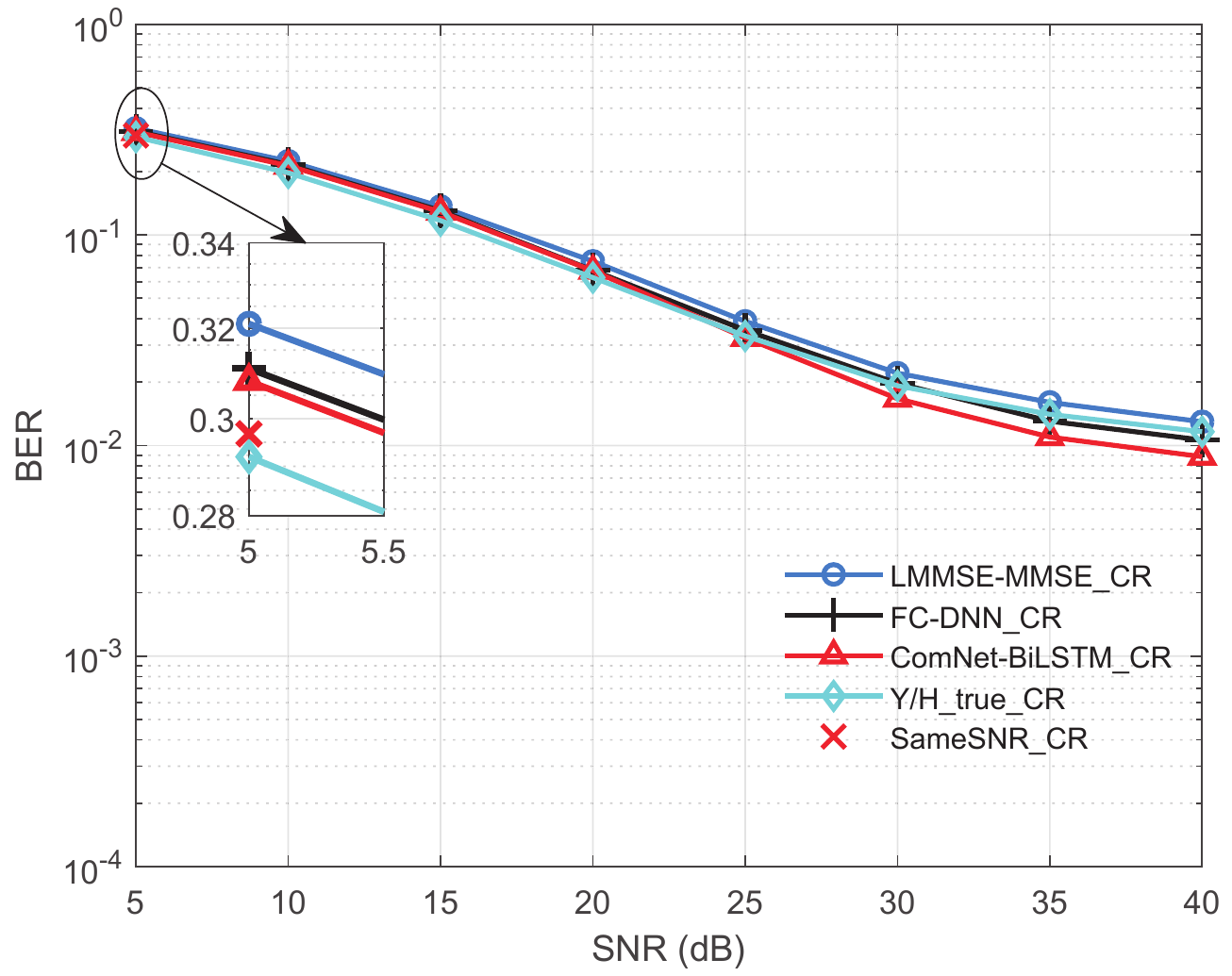}}
	\caption{BER curves of ComNet and competing methods under nonlinear cases: (a) CP removal case marked as ``\_CP'', and (b) clipping case denoted as ``\_CR''.}
	\label{Clipping}
\end{figure}


\subsubsection{CP Removal}

The CP is introduced into the OFDM system to mitigate the inter-symbol interference (ISI) caused by multipath channels, which also decreases transmission efficiency and increases energy costs. Fig. \ref{Clipping}(a) compares the BER performance of ComNet-BiLSTM with the other methods when the CP is omitted. In this case, the traditional LMMSE-MMSE method becomes saturated when SNR equals 20dB while the DL-based approaches including FC-DNN and ComNet-BiLSTM perform better in resolving ISI. In particular, ComNet-BiLSTM has about 50\% BER of FC-DNN when SNR is over 25dB, which suggests that ComNet-BiLSTM has the ability to recover transmit symbols more accurately than the other approaches for OFDM systems without CP. This ability benefits from the BiLSTM recurrent neural network that is designed to utilize the inner-relationship of ISI of the sequential data.


\subsubsection{Clipping}
One of the most detrimental characteristics in OFDM is the high peak-to-average power ratio (PAPR) \cite{Cho:2010:MWC:1951613}. A common method to reduce PAPR is the clipping operation, which is applied to the time-domain transmitted signal as in \cite{8052521} and at the same time causes nonlinear distortion on the signal. Fig. \ref{Clipping}(b) shows the BER curves of ComNet-BiLSTM and competing methods with nonlinear distortion of clipping {with the clipping ratio of 1.6}. From the figure, the ComNet-BiLSTM obtains the lowest BER among all methods.

\vspace{-0.4cm}
\subsection{Performance Analysis}
\subsubsection{Robustness of SNR Mismatching}
The abovementioned results are obtained by training the ComNet receiver offline under $\rm{SNR}=40~dB$ while deploying it online under arbitrary $\rm SNRs$, which are SNR mismatched results, whereas ``SameSNR'' markers in Figs. \ref{MSE}-\ref{Clipping} represent SNR matched results. The difference of SNR mismatched and matched results indicates that the SNR mismatching causes an MSE loss of approximately 3 dB when $\rm{SNR}=5~dB$. Nonetheless, the SNR mismatching leads to slight BER performance loss, which suggests the robustness of the ComNet receiver against SNR mismatching when recovering binary symbols.


\begin{table}[!t]
	\centering
	\caption{Complexity analysis for ComNet and competing methods}
	\label{table_example}
	\footnotesize
	\begin{tabular}{>{\sf }lllll}    %
		\toprule
		& \# of FLOPs  & Memory & Intensity & Time \\
		\midrule
		\rowcolor{mygray}
		ComNet-BiLSTM    & 10.40M  & 2.40MBytes & 4.33 & 7.2e-6s \\	
		ComNet-FC    & 0.37M  & 1.22MBytes & 0.30 & 1.2e-6s \\
		\rowcolor{mygray}
		FC-DNN         & 4.62M  & 9.30MBytes & 0.49 & 1.2e-6s \\
		{LMMSE-MMSE}   & {1.6M}  & {$-$} & {$-$} & {$-$} \\
		\bottomrule
	\end{tabular}
\end{table}

\subsubsection{Application Complexity}
TABLE \uppercase\expandafter{\romannumeral1} compares the complexities of the receivers in terms of the amount of floating-point multiplication-adds (FLOPs), memory usage, computational intensity, and time consumption required to complete a single-forward pass of one OFDM symbol. To achieve better BER performance than FC-DNN within the same time period, ComNet-BiLSTM needs more than twice FLOPs than FC-DNN with approximately one-fourth memory whereas ComNet-FC only needs 0.37 million FLOPs and 1.22 MBytes of memory. {Compared with traditional methods, the ComNet-FC consumes fewer FLOPs than LMMSE-MMSE because the parameters in ComNet receiver are settled once the training process is completed, whereas the weighted matrix of LMMSE channel estimation has to be recomputed along with the varying of channel states.}



\vspace{-0.4cm}
\section{Conclusion}
In this article, we demonstrate the benefits of the proposed ComNet receiver architecture to recover the transmit data in the OFDM system with linear and nonlinear distortions. Although the coarse-to-fine idea in the ComNet is intuitive, it offers deeper insights into its implications. From a communication viewpoint, the nonlinear activation functions in DL neural network introduce nonlinearity into the SD module, which constitutes a nonlinear signal detector. From the perspective of model-driven DL assisted by communication intelligence, the useful novel features can be created manually. Moreover, these novel features can accelerate the training process, which then results in efficient deployment performance. The idea of combining DL with expert knowledge in the ComNet receiver inspires the future work of applying model-driven DLs to the physical layers in wireless communications.

\vspace{-0.3cm}
\bibliographystyle{IEEEtran}
\bibliography{IEEEabrv,bibtex}

%





\end{document}